# Electromechanical Probing of Ionic Currents in Energy Storage Materials

A.N. Morozovska,[1,*] E.A. Eliseev,[2] and S.V. Kalinin[3,†]

[1] Institute of Semiconductor Physics, National Academy of Science of Ukraine,
41, pr. Nauki, 03028 Kiev, Ukraine

[2] Institute for Problems of Materials Science, National Academy of Science of Ukraine,
3, Krjijanovskogo, 03142 Kiev, Ukraine

[3] The Center for Nanophase Materials Sciences, Oak Ridge National Laboratory,
Oak Ridge, TN 37922

The electrochemical processes in energy storage materials are generally linked with changes of molar volume of the host compound. Here, the frequency dependent strain response of 1D electrochemically active system to periodic electric bias is analyzed. The sensitivity and resolution of electrochemical strain measurements are compared to the current-based electrochemical impedance spectroscopy. The resolution and detection limits of interferometric and atomic force microscopy based systems for probing electrochemical reactions on the nanoscale are analyzed.

[*] morozo@i.com.ua
[†] sergei2@ornl.gov

Solid-state energy storage systems based on intercalation and reconstitution chemistries are key components of multiple energy technologies.[1,2] A number of electrochemical methods has been developed to probe mechanisms of electrochemical transformations.[3,4] Among these, electrochemical impedance spectroscopy (EIS) has gained broad popularity due to the capability to decouple individual processes based on frequency dispersion of the response.[5] However, interpretation of the impedance data for complex battery devices is often not straightforward.[5] The presence of stray impedances in the measurement circuit further complicates attribution of equivalent circuit elements to the elementary mechanisms and severely limits local SPM-based measurements.[6,7]

We note that the fundamental difference between the electronic (e.g. charge-transfer resistance, double layer and stray capacitances) and ionic (Warburg impedance) elements of the equivalent circuit is that ionic transport is strongly coupled to the changes in molar volume of the material, and hence the strain state of the system. Hence, detection of the dynamic strain response in parallel with EIS measurements can provide additional information on the battery operation, and also allow extending these measurements to the nanoscale. Here, we develop the analytical theory of this electrochemical strain spectroscopy for the 1D system, and analyze the detection and sensitivity limits.

The schematic of the system is shown in Fig. 1a. Following Ho et al [8], we consider electroactive Li-containing film of thickness, $h$, sandwiched between the electronically conducting but ionically blocking bottom electrode and the ionically conducting upper electrode. The upper electrode is considered to be mechanically free (e.g. ultra-thin, or liquid, or soft polymer), so that its motion does not affect significantly on the mechanical displacement of the electrolyte film surface. The corresponding equivalent circuit is shown in Fig. 1b. Here, $Z_W$ is Warburg impedance describing the frequency response of diffusing species.[8] Note that only the Warburg impedance is electromechanically (or "strain") active.

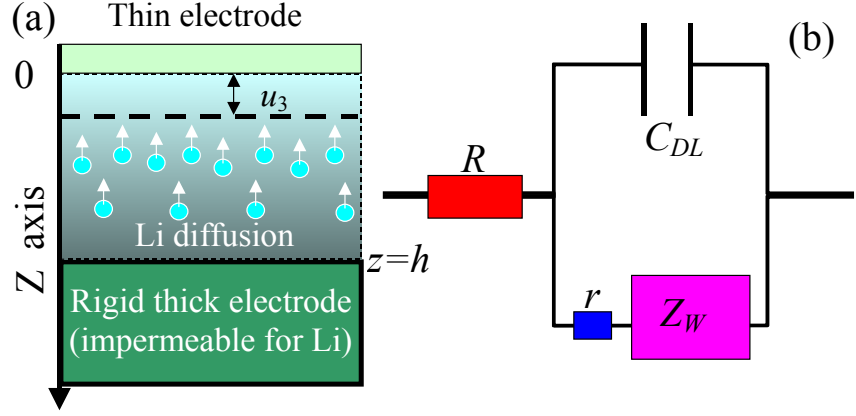

**Fig. 1.** (a) Schematic of calculated system. $u_3$ defines surface displacement for fixed back interface. (b) The equivalent circuit of the battery cell. $R$ is the ohmic resistance of the electrodes and interfacial layers, $C_{DL}$ is the double layer capacitance of the electroactive interfaces, $r$ is the charge transfer resistance, and $Z_W$ is a Warburg impedance describing the diffusion of the electroactive species. (Adapted from [8])

To establish the dynamic current and electromechanical responses, we solve the coupled strain-diffusion equations. The equations of state for isotropic elastic media relate the mole fraction variation $\delta X$, mechanical stress tensor $\sigma_{ij}$ and elastic strain $u_{ij}$ as $u_{ij} = \beta_{ij}\delta X + s_{ijkl}\sigma_{kl}$.[9] Here the excess concentration $\delta X = \delta C/C_0$, where $C_0$ is the maximum Li concentration possible by stoichiometry in the film material, $\delta C$ is the small variation ($|\delta C| \ll C_0$) of the concentration $C = C_0 + \delta C$. The $s_{ijkl}$ is the tensor of elastic compliances, and $\beta_{ij}$ is the dimensionless Vegard tensor. We further restrict the analysis to the transversally isotropic tensor $\beta_{ij} = \delta_{ij}\beta_{ii}$ ($\delta_{ij}$ is the Kroneker delta symbol) with $\beta_{11} \approx \beta_{22} \neq \beta_{33}$. Mechanical boundary conditions are defined on the mechanically free interface, $z = 0$, where the normal stress is absent, $\sigma_{3i}(0,t) = 0$, and on clamped interface $z = h$, where the displacement $u_i$ is fixed, $u_i(h,t) = 0$. The equation of mechanical equilibrium $\partial \sigma_{ij}(z,t)/\partial x_j = 0$ and compatibility conditions $\text{inc}(i,j,\hat{u}) = e_{ikl}e_{jmn}u_{ln,km} = 0$ lead to the solution of the 1D-problem as $\sigma_{11}(z,t) = \sigma_{22}(z,t) \approx -\beta_{11}\delta X(z,t)/(s_{11}+s_{12})$ and all other $\sigma_{ij} = 0$. The corresponding strains are $u_{33}(z,t) = (\beta_{33} - 2s_{12}\beta_{11}/(s_{11}+s_{12}))\delta X(z,t)$ and all other components being zero. Mechanical displacement depth distribution is then:

$$u_3(z,t) = \int_h^z dz' \left(\beta_{33} - \frac{2s_{12}\beta_{11}}{s_{11}+s_{12}}\right)\delta X(z',t). \tag{1}$$

The Li-ion dynamics obeys the conservation equation $\partial \delta C/\partial t = -\text{div} J$, where the ionic flux $J = -D(\nabla \delta C - (RT)^{-1}\nabla(\beta_{ij}\sigma_{ij}))$ is considered in the linear on $\delta C \sim \delta X$ approximation neglecting the electromigration and the term $\sim \delta X \nabla(\beta_{ij}\sigma_{ij})$, since the stress $\sigma_{ij} \sim \delta X$.[10, 11]

In the 1D case and assuming that $D$ is independent of ionic concentration (for small driving voltages), the coupled diffusion-strain equation can be rewritten in the form of the standard diffusion equation:

$$\frac{\partial}{\partial t}\delta X(z,t) = D_R \frac{d^2}{dz^2}\delta X(z,t). \qquad (2)$$

However, the effective diffusion coefficient is renormalized due to strain coupling as $D_R \approx D\left(1 + 2\beta_{11}^2 (C_0 RT)^{-1}(s_{11}+s_{12})^{-1}\right)$, $R = 8.31$ J/mol·K is the universal gas constant, $T$ is the absolute temperature. For isotropic elastic media $s_{11} + s_{12} = (1-\nu)/Y$, where $Y$ is the Young modulus and $\nu$ is the Poisson coefficient. The boundary conditions to Eq. (2) are $d\delta X(h,t)/dz = 0$ for an Li-ions blocking electrode located at $z = h$ and chemical composition changes driven by applied bias at the thin mechanically free electrode $z = 0$: $\delta X(0,t) = x_m(t)$. Here the variation $x_m(t)$ is proportional to the small periodic electric bias $\delta V_e(t) = V_0 \sin(\omega t)$ applied between the electrode located at $z = h$ and the electrode located at $z = 0$, e.g. $x_m(t) = x_0 \sin(\omega t)$ and the constant $x_0 \approx V_0 (dV_e/dX)^{-1}_{X=const} = V_0 RT/(\gamma_A Z_c F X)$, where $\gamma_A$ is the activity coefficient, $Z_c$ is the ionic charge in the units of electron charge $e$, $F$ is Faraday constant.[8]

Thus steady state concentration $\delta X(z,\omega)$ and electric current density $j(\omega) \sim eZ_c D_R d\delta C(0)/dz$ spectra are calculated as

$$\delta X(z,\omega) = \frac{\cosh\left((1+i)\sqrt{\omega/2D_R}(h-z)\right)}{\cosh\left((1+i)\sqrt{\omega/2D_R}\,h\right)} x_0, \qquad (3)$$

$$j(\omega) = eZ_c C_0 D_R (1+i)\sqrt{\frac{\omega}{2D_R}} \tanh\left((1+i)\sqrt{\frac{\omega}{2D_R}}\,h\right) x_0. \qquad (4)$$

The specific Warburg impedance $Z_W$ is inversely proportional to $j$, namely $Z_W(\omega) \cong V_0/j(\omega)S$ ($S$ is the surface area of the electroactive element). The steady state displacement field spectra $u_3(z,\omega)$ was derived from Eq.(2) and (3) as:

$$u_3(z,\omega) = \left(\beta_{33} - \frac{2s_{12}\beta_{11}}{s_{11}+s_{12}}\right) \frac{-\sinh\left((1+i)\sqrt{\omega/2D_R}(h-z)\right)}{(1+i)\sqrt{\omega/2D_R}\cdot\cosh\left((1+i)\sqrt{\omega/2D_R}\,h\right)} x_0. \qquad (5)$$

Note that for isotropic elastic media $-s_{12}/(s_{11}+s_{12}) = \nu/(1-\nu)$.

Eq. (5) predicts two frequency regimes for the strain response. At high frequencies, both the strain response and Warburg impedance scale as $1/\sqrt{\omega}$. At low frequencies the strain response is real and constant, while the impedance is imaginary and diverges as capacitance impedance $\sim 1/\omega$, as summarized in Table 1.

**Table 1**.

| Frequency regime | $u_3(0,\omega)$ | $Z_W(\omega) \cong V_0/j(\omega)S$ |
|---|---|---|
| $\sqrt{h^2\omega/2D_R} \ll 1$ | $-\left(\beta_{33} - \dfrac{2s_{12}\beta_{11}}{s_{11}+s_{12}}\right)x_0 h$ | $\dfrac{-i \cdot V_0}{eZ_c C_0 S \cdot x_0 h\omega}$ |
| $\sqrt{h^2\omega/2D_R} \gg 1$ | $-\left(\beta_{33} - \dfrac{2s_{12}\beta_{11}}{s_{11}+s_{12}}\right)\dfrac{x_0(1-i)}{\sqrt{2\omega/D_R}}$ | $\dfrac{(1-i) \cdot V_0}{eZ_c C_0 D_R S \cdot x_0 \sqrt{2\omega D_R}}$ |

Diffusion coefficients $D$ and $D_R$, elastic moduli tensor $\beta_{ii}$ and $C_0$ values used in calculations for $LiCoO_2$, $LiMn_2O_4$ and $LiC_6$ materials are listed in the Table 2. It is seen that the strain renormalization can be essential in some cases.

**Table 2**. Material parameters.

| Composition | $D_R$ (cm$^2$/s) | $D^*$) (cm$^2$/s) | $\nu$ | $Y$ (GPa) | $\beta_{22} = \beta_{11}$ | $\beta_{33}$ | $C_0$ (kmol/m$^3$) | Refs. |
|---|---|---|---|---|---|---|---|---|
| $LiCoO_2$ | $2.5\times10^{-12}$ | $2.4\times10^{-12}$ | 0.27 | 100 | <0.01 | 0.078 | 51.6 | [12, 13, 14] |
| $LiMn_2O_4$ | $7.08\times10^{-11}$ | $3.83\times10^{-11}$ | 0.33 | 25 | 0.027 | 0.027 | 22.9 | [15] |
| $LiC_6$ | $1.0\times10^{-11}$ across layer | $0.8\times10^{-11}$ | 0.3 | 52 | 0.012 | 0.104 | 30.6 | [15] |

*) $D$ values were provided by Edwin Garcia

Frequency dependence of displacement $\tilde{u}_3$ and inverse current density $1/j$ (proportional to $Z_W$) for several Li-containing materials $LiCoO_2$, $LiC_6$ and $LiMn_2O_4$ are shown in Fig. 2. Note the close similarity between the strain and impedance behavior and cross-over at the frequencies corresponding to the transition from finite to infinite diffusion in both cases. The corresponding Cole-Cole plots are shown in Fig. 3.

Displacement $\tilde{u}_3$ dependence on the external field frequency $f$ calculated for several film thicknesses is shown in Figure 4. Note the evolution of maximal response amplitude and the cross-over frequency with film thickness.

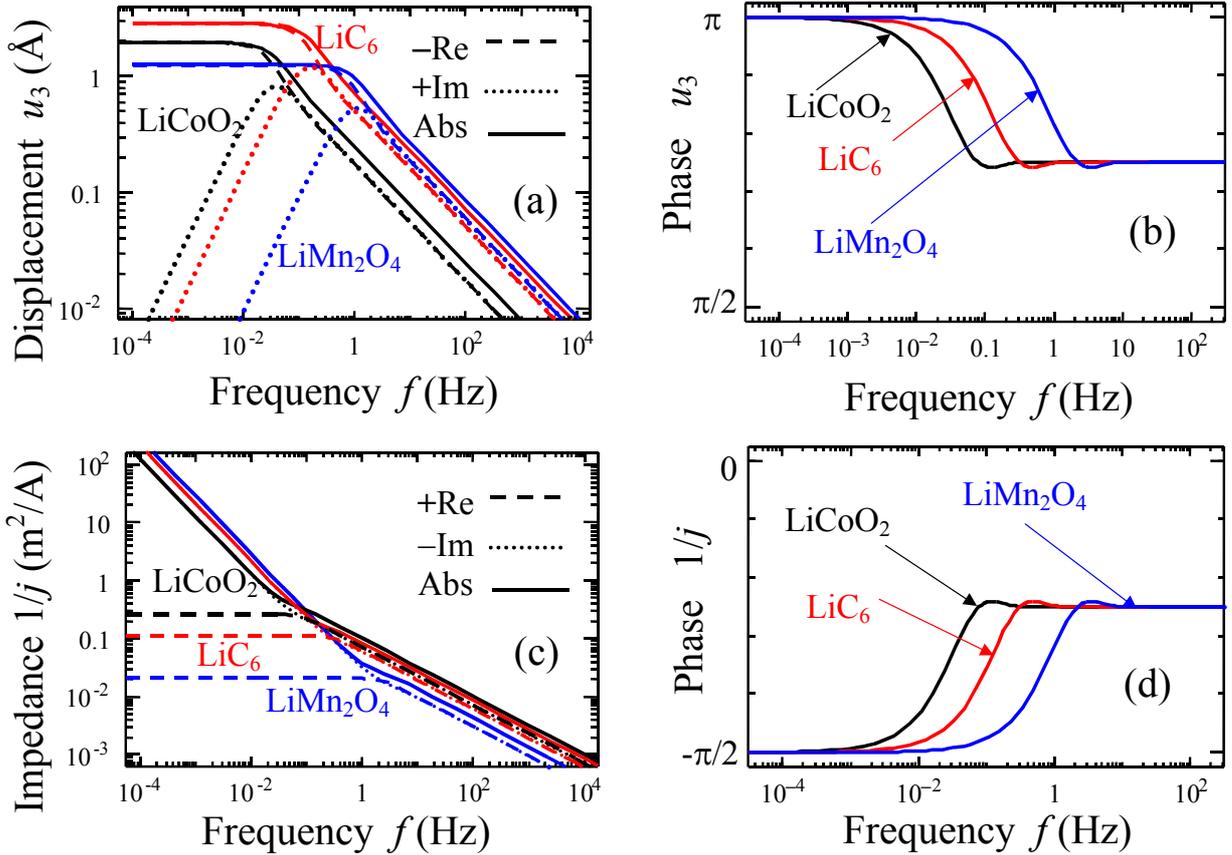

**Fig. 2.** The real (Re), imaginary (-Im) parts, absolute value (Abs) and phase of displacement $\tilde{u}_3$ (a,b) and inverse current density $1/j$ (c,d) vs. external field frequency $f$ calculated for Li-containing materials $LiCoO_2$, $LiC_6$ and $LiMn_2O_4$. Film thickness $h = 50$ nm, fraction $x_0 = 0.05$.

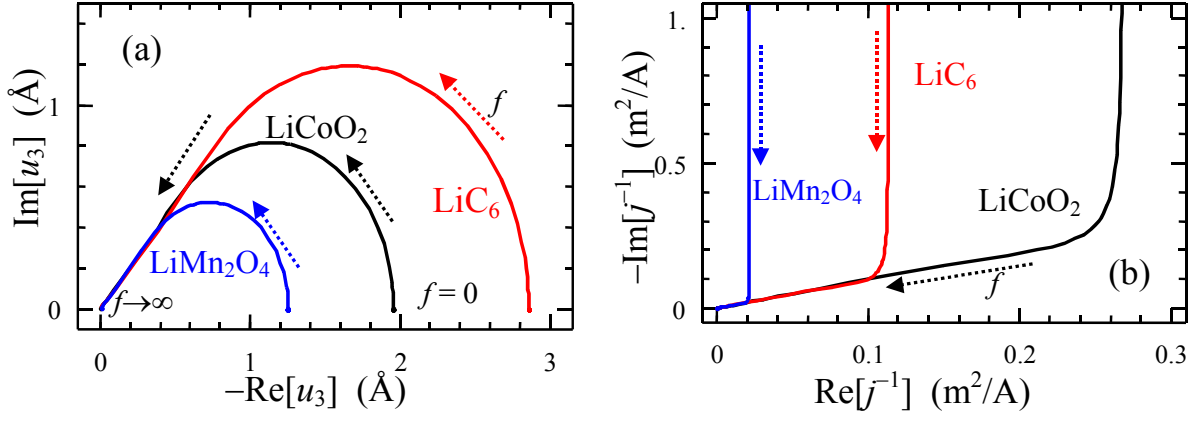

**Fig. 3.** Dependences of Im[$u_3$] vs. Re[$u_3$] (a) and Im[$1/j$] vs. Re[$1/j$] (b) calculated for Li-containing materials $LiCoO_2$, $LiC_6$ and $LiMn_2O_4$. The frequency $f$ changes from low to high values along the arrow. Film thickness $h = 50$ nm, fraction $x_0 = 0.05$.

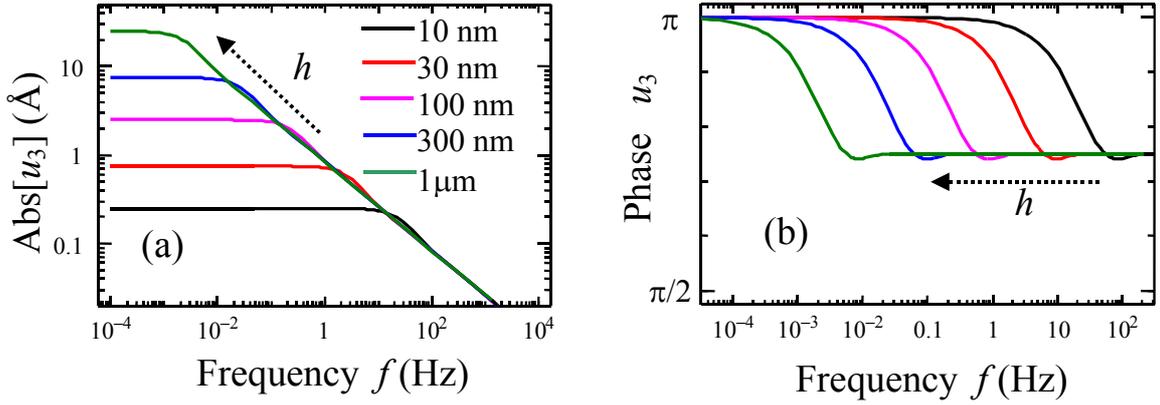

**Fig. 4.** The absolute value (a) and phase (b) of displacement $\tilde{u}_3$ vs. external field frequency $f$ calculated for $LiMn_2O_4$. Different curves correspond to the film thickness $h = 10, 30, 100, 300, 10^3$ nm, fraction $x_0 = 0.05$.

The analysis in Eq. (3-5) and Figs. 2,3 suggest the close similarity between the current and strain responses of the Warburg element, as expected for the case of linear strain-concentration relationship. However, the significant differences arise on considering the detection limits and the contributions of the other elements of equivalent circuit to the measured signal. For current-based EIS measurements, all elements of equivalent circuit contribute to measured response. In the strain measurements, only the response of the Warburg element is probed, and the latter can thus be separated from all parallel elements (while the serial elements will determine the effective bias acting on the element).

The sensitivity and minimal probed volume of the EIS measurements is governed by multiple instrumental factors. Considering the limits imposed by the (frequency independent) ~1 GΩ sensitivity of a typical commercial tool and ~40 pF cable capacitances, the detection limit for current detection (for 10 mV driving voltage at cross-over frequency) corresponds to the device size of the order of 10 – 100 μm and the volume of $10^{10} - 10^{14}$ nm$^3$. In comparison, for double-beam interferometer systems reported detection limit can be ~0.3 pm at 8 kHz.[16] From Figs. 3-4, this enables measurements (for 5% changes in concentration) down to ~10 nm film thicknesses. Assuming complete charge-discharge behavior, films as thin as ~1 nm can be probed. The device size is limited by the beam spot size, but can be as small as ~20 microns, corresponding to effective material volume of $10^8$ nm$^3$. For atomic force microscopy methods, the vertical sensitivity of ~10 pm at 1 kHz can be achieved, enabling measurements of 10-100 nm films. However, the lateral size can be as small as 10-30 nm, corresponding to the probing volumes of $10^3 - 10^4$ nm, which is ~6 orders of magnitude below the EIS limits. Note that multiple other methods for detecting strains in small volumes, e.g. using cantilever systems or surface force apparatus, can be considered.

To summarize, an approach for studies of electrochemical processes based on the frequency dependence of strain response is discussed and compared with classical electrochemical impedance measurements. The frequency dependence of response is derived. The method is projected to allow probing of electrochemical behaviors in $10^6$ smaller volumes then currently possible.

This material is based upon work supported as part of the Fluid Interface Reactions, Structures and Transport (FIRST) Center, an Energy Frontier Research Center funded by the U.S. Department of Energy, Office of Science, Office of Basic Energy Sciences under Award Number ERKCC61 (S.V.K.).

**Appendix A. Elastic problem solution (available in ArXiv only)**

For the quasi-static case, the equations of state $\beta_{ij}\delta X + s_{ijkl}\sigma_{kl} = u_{ij}$ for elastically isotropic media can be rewritten as:[17]

$$u_{11} = s_{11}\sigma_{11} + s_{12}(\sigma_{22} + \sigma_{33}) + \beta_{11}\delta X, \quad u_{22} = s_{11}\sigma_{22} + s_{12}(\sigma_{11} + \sigma_{33}) + \beta_{22}\delta X,$$
$$u_{33} = s_{11}\sigma_{33} + s_{12}(\sigma_{22} + \sigma_{11}) + \beta_{33}\delta X, \quad (A.1)$$
$$u_{12} = (s_{11} - s_{12})\sigma_{12}, \quad u_{13} = (s_{11} - s_{12})\sigma_{13}, \quad u_{23} = (s_{11} - s_{12})\sigma_{23}.$$

Here $\delta C$ is the concentration redistribution.

Then equation of mechanical equilibrium $\partial \sigma_{ij}(z,t)/\partial x_j = 0$ and compatibility conditions $\text{inc}(i,j,\hat{u}) = e_{ikl} e_{jmn} u_{ln,km} = 0$ lead to equations

$$0 = u_{11} = s_{11}\sigma_{11} + s_{12}\sigma_{22} + \beta_{11}\delta X, \quad 0 = u_{22} = s_{11}\sigma_{22} + s_{12}\sigma_{11} + \beta_{22}\delta X,$$
$$0 = u_{22} = s_{11}\sigma_{22} + s_{12}\sigma_{11} + \beta_{22}\delta X, \quad (A.2)$$
$$u_{12} = (s_{11} - s_{12})\sigma_{12} = 0, \quad u_{13} = (s_{11} - s_{12})\sigma_{13} = 0, \quad u_{23} = (s_{11} - s_{12})\sigma_{23} = 0.$$

The solution for nonzero components is

$$\sigma_{11} = \left(\frac{s_{12}\beta_{22} - s_{11}\beta_{11}}{s_{11}^2 - s_{12}^2}\right)\delta X \approx -\frac{\beta_{11}\delta X}{s_{11} + s_{12}},$$
$$\sigma_{22} = \left(\frac{s_{12}\beta_{11} - s_{11}\beta_{22}}{s_{11}^2 - s_{12}^2}\right)\delta X \approx -\frac{\beta_{11}\delta X}{s_{11} + s_{12}}, \quad (A.3)$$
$$u_{33} = \left(\beta_{33} - \frac{s_{12}(\beta_{11} + \beta_{22})}{s_{11} + s_{12}}\right)\delta X \approx \left(\beta_{33} - \frac{2s_{12}\beta_{11}}{s_{11} + s_{12}}\right)\delta X.$$

The approximate equalities correspond to the almost transversally isotropic Vegard tensor $\beta_{11} \approx \beta_{22} \neq \beta_{33}$.

**Appendix B. Diffusion problem solution (available in ArXiv only)**

Considered Li-ion dynamics obeys the conservation equation $\partial \delta C/\partial t = -\text{div} J$, where the ionic flux $J = -D(\nabla \delta C - (RT)^{-1}\nabla(\beta_{ij}\sigma_{ij}))$ is considered in the linear on $\delta C \sim \delta X$ approximation, i.e. keeping in mind that the stress $\sigma_{ij} \sim \delta X$ and so the term $\sim (RT)^{-1}\delta X \nabla(\beta_{ij}\sigma_{ij})$ could be neglected ($R$ = 8.31 J/mol·K is the universal gas constant, $T$ is the absolute temperature). [18]

Substituting here $\sigma_{11} = -\frac{\beta_{11}\delta X}{s_{11} + s_{12}}, \quad \sigma_{22} = -\frac{\beta_{11}\delta X}{s_{11} + s_{12}}$ from Eqs.(A.3) we derived that $J = -D\left(\nabla \delta X + \frac{\beta_{11}^2 \nabla \delta X}{(s_{11} + s_{12})RT}\right)C_0 = -D\left(1 + \frac{\beta_{11}^2}{(s_{11} + s_{12})RT}\right)C_0\nabla\delta X = -C_0 D_R \nabla \delta X$, where the effective diffusion coefficient is renormalized due to strain coupling as $D_R \approx D\left(1 + 2\beta_{11}^2 C_0 (RT)^{-1}(s_{11} + s_{12})^{-1}\right)$ for $\beta_{11} = \beta_{22}$. So, in the 1D case and assuming that $D$ is independent on ionic concentration, the coupled diffusion-strain equation can be rewritten in the form of the standard diffusion equation:

$$\frac{\partial}{\partial t}\delta X(z,t) = D_R \frac{d^2}{dz^2}\delta X(z,t). \qquad (2)$$

Not considering the transient process, Eq.(2) with boundary conditions has the form

$$\begin{cases} \dfrac{\partial}{\partial t}\delta X(z,t) = D_R \dfrac{d^2}{dz^2}\delta X(z,t), \\ \delta X(0,t) = x_m(t) = x_0 \exp(i\omega t), \\ \dfrac{d}{dz}\delta X(h,t) = 0 \end{cases} \qquad (B.1)$$

We looking for the periodic solution of the system (B.1) at frequency ω in the form $\delta X(z,t) = \delta \tilde{X}(z,\omega)\exp(i\omega t)$, we obtained

$$\begin{cases} i\omega\delta\tilde{X}(z,\omega) = D_R \dfrac{d^2}{dz^2}\delta\tilde{X}(z,\omega), \\ \delta\tilde{X}(0,\omega) = x_0, \\ \dfrac{d}{dz}\delta\tilde{X}(h,\omega) = 0 \end{cases} \qquad (B.2)$$

The solution of the system (B.2) is $\delta\tilde{X}(z,\omega) = C_1 \exp(z\sqrt{i\omega/D_R}) + C_2 \exp(-z\sqrt{i\omega/D_R})$, where the constant $C_{1,2}$ are determined from the boundary conditions, namely $C_1 + C_2 = x_0$ and $C_1 \exp(h\sqrt{i\omega/D_R}) - C_2 \exp(-h\sqrt{i\omega/D_R}) = 0$. From here the constants are

$$C_2 = \frac{x_0}{1+\exp(-2h\sqrt{i\omega/D_R})}, \quad C_1 = \frac{x_0 \exp(-2h\sqrt{i\omega/D_R})}{1+\exp(-2h\sqrt{i\omega/D_R})}.$$

Thus $\delta\tilde{X}(z,\omega) = x_0 \dfrac{\exp((z-2h)\sqrt{i\omega/D_R}) + \exp(-z\sqrt{i\omega/D_R})}{1+\exp(-2h\sqrt{i\omega/D_R})} \equiv \dfrac{\cosh(\sqrt{i\omega/D_R}(h-z))}{\cosh(\sqrt{i\omega/D_R}\,h)} x_0.$

Using that $\sqrt{i} = \pm\dfrac{1+i}{\sqrt{2}}$, concentration and electric current density spectra were calculated

$$\delta\tilde{X}(z,\omega) = \frac{\cosh((1+i)\sqrt{\omega/2D_R}(h-z))}{\cosh((1+i)\sqrt{\omega/2D_R}\,h)} x_0, \qquad (B.3)$$

$$j(\omega) = eZ_c D_R \frac{d\delta C(\omega, z=0)}{dz} = eZ_c D_R C_0 \frac{d\delta \widetilde{X}(z=0,\omega)}{dz}$$
$$= eZ_c C_0 D_R (1+i)\sqrt{\frac{\omega}{2D_R}} \tanh\left((1+i)\sqrt{\frac{\omega}{2D_R}} h\right) x_0 \quad . \tag{B.4}$$

The solution in time domain that satisfy the boundary condition $\delta X(0,t) = x_0 \sin(\omega t)$ should be constructed as $\delta X(z,t) = \frac{1}{2i}\left(\delta\widetilde{X}(z,\omega)\exp(i\omega t) - \delta\widetilde{X}(z,-\omega)\exp(-i\omega t)\right)$, namely

$$\delta X(z,t) = \frac{x_0}{2i}\left(\frac{\cosh\left(\sqrt{i\omega/D_R}(h-z)\right)\exp(i\omega t)}{\cosh\left(\sqrt{i\omega/D_R}\,h\right)} - \frac{\cosh\left(\sqrt{-i\omega/D_R}(h-z)\right)\exp(-i\omega t)}{\cosh\left(\sqrt{-i\omega/D_R}\,h\right)}\right) \tag{B.5}$$